\documentclass[12pt,preprint]{aastex}

\shorttitle{Relativistic reconnection} \shortauthors{Karlick\'y}
\begin{document}

\title{Separation of accelerated electrons and positrons in the
relativistic reconnection}

\author{Marian Karlick\'y}
\affil{Astronomical Institute of the Academy of Sciences of the Czech Republic,
 CZ-25165 Ond\v{r}ejov, Czech Republic}
 \email{karlicky@asu.cas.cz}

\begin{abstract}
We study an acceleration of electrons and positrons in the relativistic
magnetic field reconnection using a 2.5-D particle-in-cell electromagnetic
relativistic code. We consider the model with two current sheets and periodic
boundary conditions. The electrons and positrons are very effectively
accelerated during the tearing and coalescence processes of the reconnection.
We found that near the X-points of the reconnection the positions of electrons
and positrons differ. This separation process is in agreement with those
studied in the previous papers analytically or by test particle simulations. We
expect that in dependence on the magnetic field connectivity this local
separation can lead to global spatial separation of the accelerated electrons
and positrons. A similar simulation in the electron-proton plasma with the
proton-electron mass ratio $m_i/m_e$ = 16 is made.
\end{abstract}


\section{INTRODUCTION}

Magnetic reconnection is the key process in conversion of the magnetic field
energy into particle kinetic energy. It is well accepted that it plays a
crucial role in the Earth's magnetotail, solar flares and accretion discs
(Priest \& Forbes 2000, Drake et al. 2005, Pritchett 2006). The relativistic
reconnection in electron-positron plasmas is proposed for high energy
astrophysical phenomena including the jets from active galactic nuclei (Lesch
\& Birk 1998, Larrabee et al. 2003, Wardle et al. 1998), pulsar winds (Coroniti
1990, Michel 1994, Lyubarsky \& Kirk 2001) and models of gamma-ray bursts
(Drenkhahn 2002a,b). The relativistic reconnection and particle acceleration in
pair plasmas was studied numerically for the first time by Zenitani \& Hoshino
(2001, 2005). The effectiveness of such an acceleration and corresponding
synchrotron spectra has been computed in detail in the paper by Jaroschek et
al. (2004a,b). Bessho \& Bhattacharjee (2005) have shown that this fast
reconnection is caused by the off-diagonal components of the pressure tensor.

Recently, the RHESSI observations of the solar flare on 2002 July 23 have
revealed a separation of the gamma-ray source from any of those observed in the
hard X-ray emission. It has been interpreted as a spatial separation of
energetic electrons and protons (Share et al. 2003). Based on the analytical
and test particle approach Zharkova \& Gordovskyy (2004) have explained this
separation by the asymmetry in acceleration of electrons and protons in the
reconnecting non-neutral current sheet, see also the papers by Martens \& Young
(1990), Zhu \& Parks (1993), Litvinenko (1996).

This separation acceleration can be even more distinct in pair plasmas due to
the same mass of electrons and positrons. Therefore in this paper using the
particle-in-cell modelling we study this process in the electron-positron
plasma in detail.

\section{MODEL}

We used a 2.5-D (2D3V -- 2 spatial and 3 velocity components) fully
relativistic electromagnetic particle-in-cell code (Saito \& Sakai 2004). The
system size is $L_x \times L_y$ = 2000$\Delta$ $\times$ 600$\Delta$ = 200$d_e$
$\times$ 60$d_{e}$, where $\Delta$ (=1) is a grid size, $d_e$ = $c/\omega_{pe}$
is the electron inertial length, $c$ is the speed of light and $\omega_{pe}$ is
the plasma frequency.

Two 2-D current sheets with the guiding magnetic field B$_z$ are initiated
along the lines $y$ = 150$\Delta$ and $y$ = 450$\Delta$. The periodic boundary
conditions are used. The half-width of both the current sheets is 10$\Delta$ =
$d_e$. The initial magnetic field is (see also Karlick\'y \& B\'arta 2007)
\begin{eqnarray}
{\bf B} \equiv ({B_x, B_y, B_z}),\nonumber \\
B_x = -B_0~ \rm{for~y} < 140 \Delta ,\nonumber \\ B_x = (y-150) B_0/10~{\rm
for}~ 140 \Delta \leq~{\rm y} \leq~160 \Delta, \nonumber \\ B_x = B_0~
\rm{for}~ 160
\Delta < y < 440 \Delta , \nonumber \\
B_x = -(y-450) B_0/10~{\rm for}~ 440  \Delta \leq~{\rm y} \leq~460 \Delta, \nonumber \\
B_x = - B_0 ~ \rm{for~y} > 460 \Delta ,\nonumber \\
B_y = 0, B_z =B_0. \nonumber
\end{eqnarray}

We consider the electron-positron plasma. In each numerical cell located out of
the current sheet we initiated n$_0$ = 60 electrons and n$_0$ = 60 positrons.
In this region out of the current sheet we define the plasma frequency for the
time unit $\omega_{pe}^{-1}$. The time step in computations is $\omega_{pe}
\Delta t$ = 0.05.  The total amount of particles in the model is 172 millions.
The initial number density is enhanced in the current sheets just to keep the
pressure equilibrium in the current sheet. The particle distribution is taken
as by Zenitani \& Hoshino (2001) in the form of f $\sim \exp
{-m[u_x^2+u_y^2+(u_z-U)^2]/2T}$, where the velocity $u$ is related to the
particle velocity $v$ as $u = v \gamma = v/[1-(v/c)^2]^{1/2}$, $U$ is the drift
velocity, $m$ is the electron rest mass, $T$ is the plasma temperature and $c$
is the speed of light. The mean initial thermal energy of electrons and
positrons is taken the same as 0.45 $mc^2$. We neglect any collisions, pair
production, and pair annihilation of pair plasmas.

Due to our interest about the reconnection processes in the relativistic plasma
with high magnetic field we consider cases with low-$\beta$ plasmas. The plasma
beta parameter and the ratio of the electron-cyclotron and electron-plasma
frequencies in the region out of the current sheets are chosen as $\beta$=0.11,
$\omega_{ce}/\omega_{pe}$ = 4 (Case I) and $\beta$=0.5,
$\omega_{ce}/\omega_{pe}$ = 1.9 (Case II). For comparison one run was made for
the parameters as in Case I, but without the guiding magnetic field, i.e. $B_z
= 0$.

Furthermore the same processes are modelled in the electron-proton plasma (Case
III) with the proton-electron mass ratio $m_i/m_e$ = 16. The proton and
electron temperature is taken the same $T_i=T_e$. The parameters are $\beta$ =
0.11 and $\omega_{ce}/\omega_{pe}$ = 4. The mean initial thermal energy is 0.45
$mc^2$.

All computations were performed on the parallel computer OCAS (Ond\v{r}ejov
Cluster for Astrophysical Simulations), see http://wave.asu.cas.cz/ocas.

\section{RESULTS}

Due to the tearing mode instability the current sheet tears into O-type islands
(plasmoids) which later on coalesce into larger ones. During these processes
both the electrons and positrons are accelerated. Figure 1 shows an evolution
of the electron distribution function $f(E)$ (for Case I) in dependence on the
energy $E$ in the whole numerical plane (upper part of Figure 1) as well as in
one selected location (bottom part of Figure 1, for the location see Figure 4),
where much harder spectrum can be seen. While the spectral index at
$\omega_{pe} t$ = 600 in the whole plane is -3.3, in the selected location is
about -1.9. There are other such places, especially at the locations where
high-energy electrons and positrons are produced, see Figure 4. This
acceleration process is very efficient and looks to have a 2-step character in
time as shown in Figure 2, where an evolution of the number of accelerated
electrons in the ratio to the total number of electrons (in percents) with the
energy $E/m^2 = \gamma
>$ 4 is shown. An analysis shows that the first step of the acceleration (up to
about $\omega_{pe} t$ = 400) is connected with tearing processes and in the
second step (after $\omega_{pe} t$ = 400) the main process is the coalescence
of plasmoids. The acceleration with the higher value of
$\omega_{ce}/\omega_{pe}$ = 4 (full line) (Case I) is more efficient than that
with $\omega_{ce}/\omega_{pe}$ = 1.9 (dashed line) (Case II) (Figure 2).

Similarly as in previous studies (Drake et al. 2005, Pritchett 2006, Jaroschek
et al. 2004a,b) the electrons and positrons are accelerated in the electric
field near the X-points formed during the tearing and coalescence processes.
But we found that during this acceleration process (Figures 3 and 4) the
electrons and positrons are moving into different locations around the O-type
magnetic structures (plasmoids) and thus they are spatially separated (e.g. see
the region around $x$ = 500$\Delta$ and $y$ = 450$\Delta$ in Figure 4). To
understand this separation process we analyzed the electric field near the
X-point of the reconnection (Figures 5 and 6). As seen here the electrons (the
asterisks) are located at the borders of the areas with the enhanced ($-E_y$
and $-E_x$) and ($+E_y$ and $+E_x$). On the other hand, the positrons are
located along the remaining two borders (see Figure 3, bottom part). The
electric component $E_z$ along the line $y$ = 150$\Delta$ is negative between
the O-type magnetic structures (Figure 6). But only near the X-point of the
reconnection this electric field ($E_z$) deviates from that of the inductive
one $-{\bf v \times B}/c$ (where ${\bf v}$ is the plasma velocity), see Figure
7, where their profiles are shown at two times ($\omega_{pe} t$ = 50 and 100).
This deviation defines the diffusion region of the reconnection. As concerns
the magnetic field in the early stage of the reconnection, the magnetic field
component $B_y$ is positive in the region $x
>$ 1000$\Delta$ and negative for $x <$ 1000$\Delta$. The structure of the magnetic
field together with the plasma velocity pattern in the regions from the X-point
to the magnetic island centers resembles to that of the collapsing magnetic
trap (Giuliani et al. 2005; Karlick\'y \& B\'arta 2006). In such a structure
the particles are also accelerated, but not separated as found here near the
X-points. This additional acceleration process is known as that in the
contracting magnetic islands (Drake et al. 2006).

Furthermore, we made the similar computations, but without the guiding magnetic
field component ($B_z= 0$). In this case no separation of electrons and
positrons was found. Also the electric field structure was different from that
presented in Figures 5 and 6.

Finally, we made similar computations also for the electron-"proton" plasma,
with the electron-proton ratio $m_i/m_e$ = 16 (Case III). Similarly as in the
previous cases, Figure 8 shows that accelerated electrons and protons move to
different positions. Comparing this Case III (dotted line in Figure 2) with
Case I (full line) the number of accelerated electrons is reduced.

\section{DISCUSSION AND CONCLUSIONS}

The present simulations show that the magnetic reconnection with the guiding
magnetic field accelerate electrons to different positions around the plasmoid
than for positrons or "protons". If the magnetic field connectivity (in the
$z$-direction) from the upper and bottom part of the plasmoid differs then the
accelerated electrons and positrons (or protons) move into quite different
locations as observed by RHESSI. The separation of particles with different
electric charge is a natural consequence of the acceleration in direct electric
field near the X-points of the magnetic field reconnection.

In agreement with Litvinenko (1996) and Zharkova \& Gordovskyy (2004) we found
that the separation process is due to a presence of the non-zero guiding
magnetic field (non-neutral current sheet). Namely, our simulations with the
zero guiding magnetic field show no such separations.

Zharkova \& Gordovskyy (2004) have shown that the separation direction (in the
present designation of the electric and magnetic field components) depends on
the sign of the term $q^3 B_y B_z E_z$, where $q$ is the electron (-e) or
positron (+e) charge (see the relation (8) in their paper). Considering the
direction of the magnetic and electric fields in our case (Figures 5 and 6) it
can be shown the separation direction found agrees to this relation. This
result agrees also to the relations presented in the paper by Litvinenko
(1996).

The similar separation process is found also for the reconnection in the
"proton"-electron plasma with the proton-electron mass ratio $m_i/m_e$ = 16.
This mass ratio is not realistic and is taken due to computer limitations.
Nevertheless, we expect that such a separation process will be confirmed by
future computations also for the real proton-electron mass ratio.

Comparing the acceleration process for the electron-positron (full line) and
electron-proton ($m_i/m_e$=16, dotted line) plasma in Figure 2, we found that
N$_{m_i/m_e=1}$/N$_{m_i/m_e=16}$ is 1.18, i.e. the number of accelerated
electrons N depends on the proton-electron mass ratio as N$_{1}$/N$_{m_i/m_e}$
$\simeq$ 1/($m_i/m_e$)$^{0.0625}$. If this relation is valid also for the real
proton-electron mass ratio then we can write N$_{1}$/N$_{1838}$ = 1.6, which
gives enough accelerated electrons also for the real electron-proton plasma.

Comparing the present modelling  with previous studies the most similar
simulation is that of Zenitani \& Hoshino (2001), especially due to initial
high thermal plasma energy. But in their model no guiding magnetic field, which
is crucial for the particle separation, is considered. Furthermore, contrary to
our start from noise level they initiate the reconnection by magnetic field
perturbation which can influence the separation process, too. The maximum
energies of accelerated electrons in both models are comparable. But in our
model the reconnection process is about three times slower than that in
Zenitani \& Hoshino (2001). We think that it is due to the initial magnetic
field perturbation.

Although the acceleration in the contracting magnetic islands does not separate
particles of opposite electric charges this process is important for global
acceleration.  But this process is even more complicated than presented in the
paper by Drake et al. (2006). Namely, not only parallel energy of particles
increases due to reflection from the ends of contracting magnetic islands (as
described by the relation (1) in Drake et al. (2006)) but also the
perpendicular energy of particles $E_{\perp}$ can increase due to the betatron
type of the acceleration, which follows from the conservation of the magnetic
moment $\mu = E_{\perp}/B$ in the region with the increasing magnetic field $B$
(see Karlick\'y \& B\'arta 2006). In our simulations the contracting
acceleration is time varying, therefore let us compare its efficiency with that
of the acceleration near the X-point at one specific time. Using the relation
(1) of Drake et al. (2006) we derived the electric field equivalent to this
process as $E_{eq}$ = (v$_x$ B$_x^2$ v m$_e$)/($\delta_x$ e B$^2$), where
2$\delta_x$ is the length of the island, $B_x$ and $B$ are the reconnecting and
total magnetic fields, $v_x$ is the contracting velocity, $v$ is the mean
electron velocity, and $m_e$ is the electron mass. For the parameters in Case I
at $\omega_{pe} t$ = 100 it gives the equivalent electric field in the
contracting magnetic island $E_{eq}$ one order of magnitude lower than the
electric field $E_z$ at the X-point region. It means that at this moment the
acceleration near the X-point dominates over that in the contracting magnetic
islands.

\acknowledgments This research was supported by the Centre for Theoretical
Astrophysics and by Grants IAA300030701 of the Academy of Sciences of the Czech
Republic. M.~K.\ thanks Prof.~J.\ I.\ Sakai and Dr.~S.\ Saito for many useful
discussions concerning numerical modelling. Author also thanks to the referee
for constructive comments improving the paper.

\clearpage


\begin{figure}
    \epsscale{0.5}
\plotone{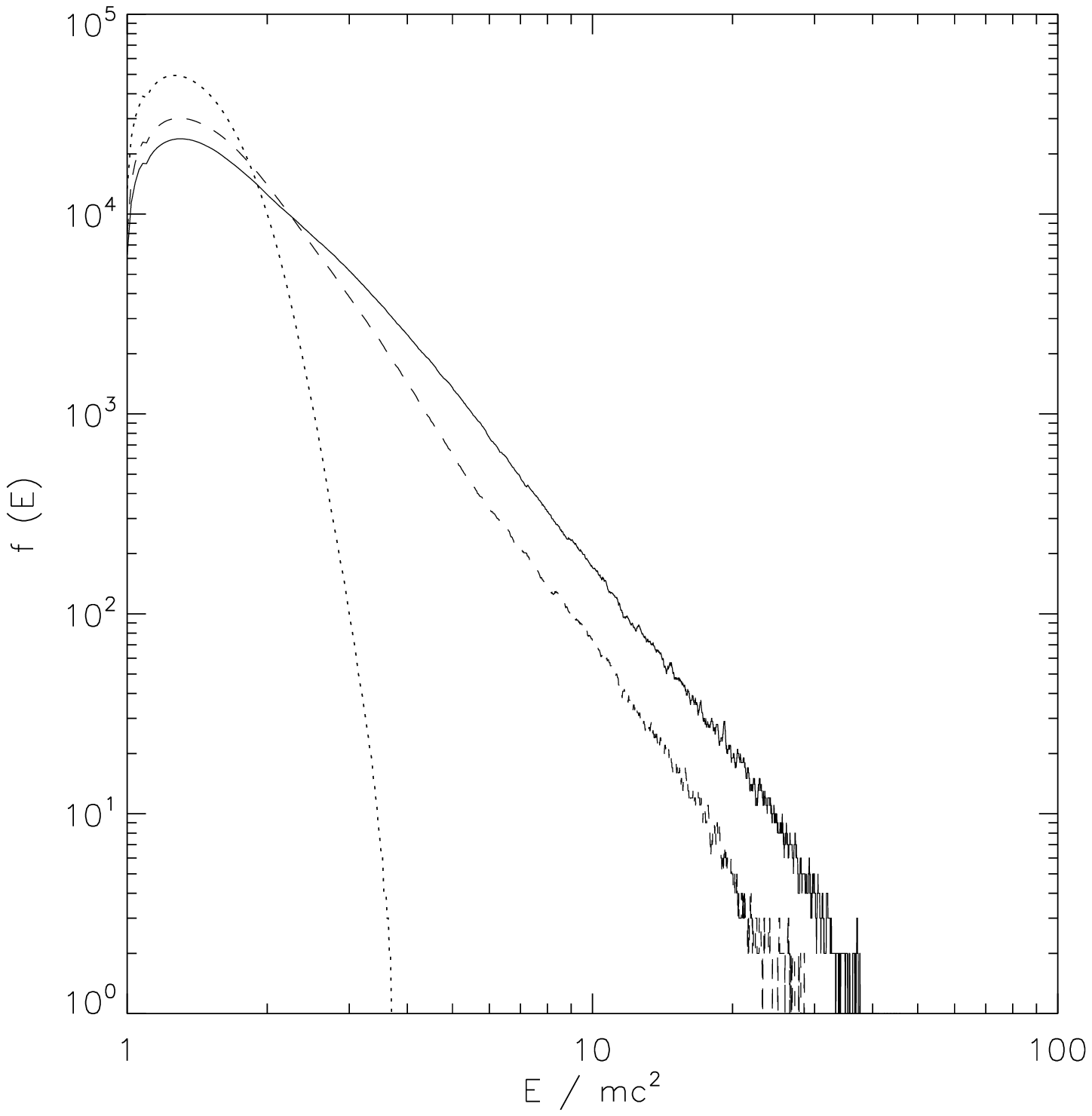} \plotone{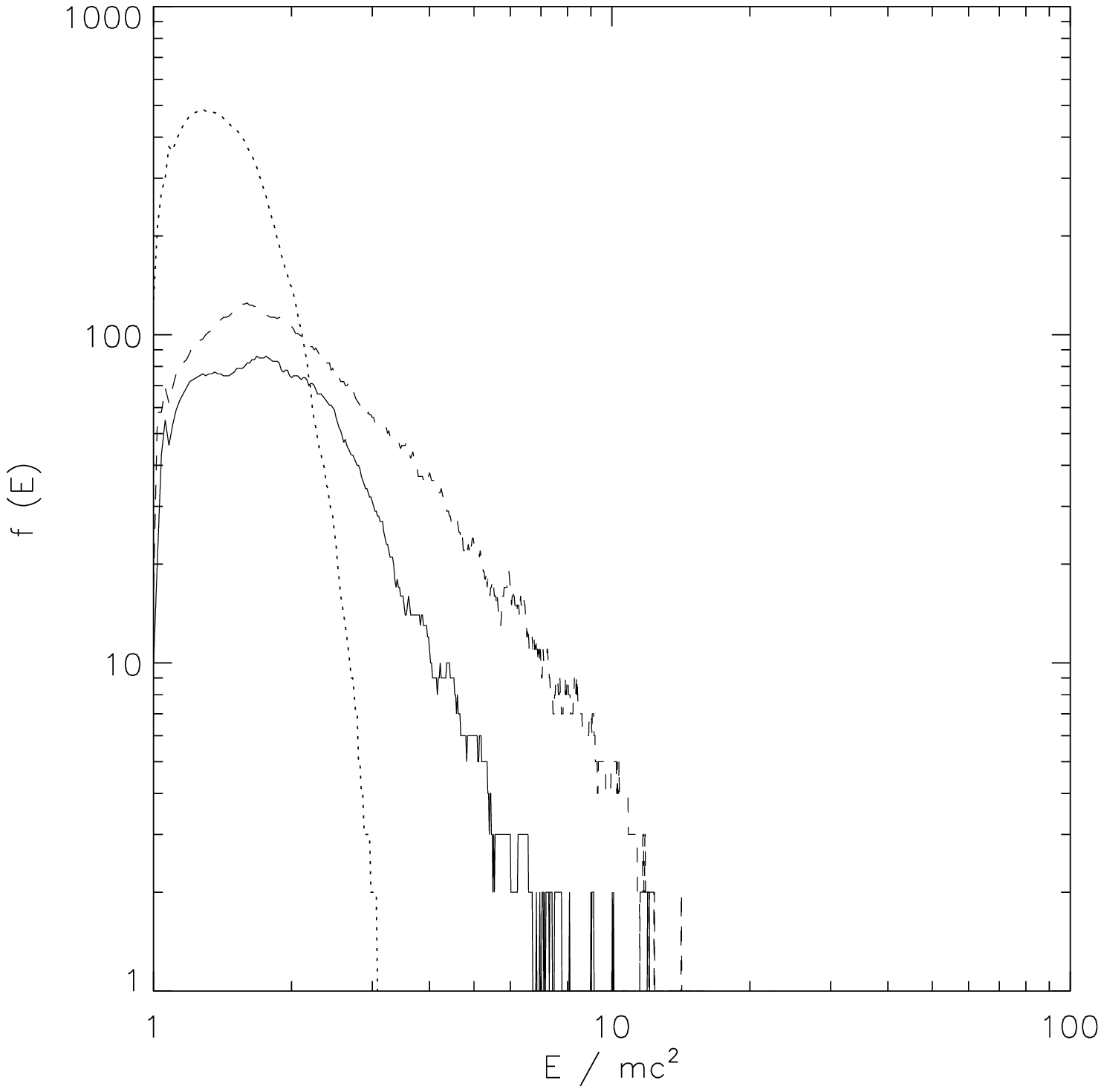}
\caption{The energy distribution of electrons at the initial state (dotted line),
    at $\omega_{pe} t$ = 600 (dashed line), and at $\omega_{pe} t$ = 1000 (full line) (Case I).
    The upper part: The distributions in the whole numerical plane. The bottom part:
    The distributions in the selected location, in the circle centered
    at $x$=550$\Delta$ and $y$=450$\Delta$ with the radius $r$ = 50$\Delta$.
    For this location, see Figure 4.}
\end{figure}

\begin{figure}
    \epsscale{1.}
\plotone{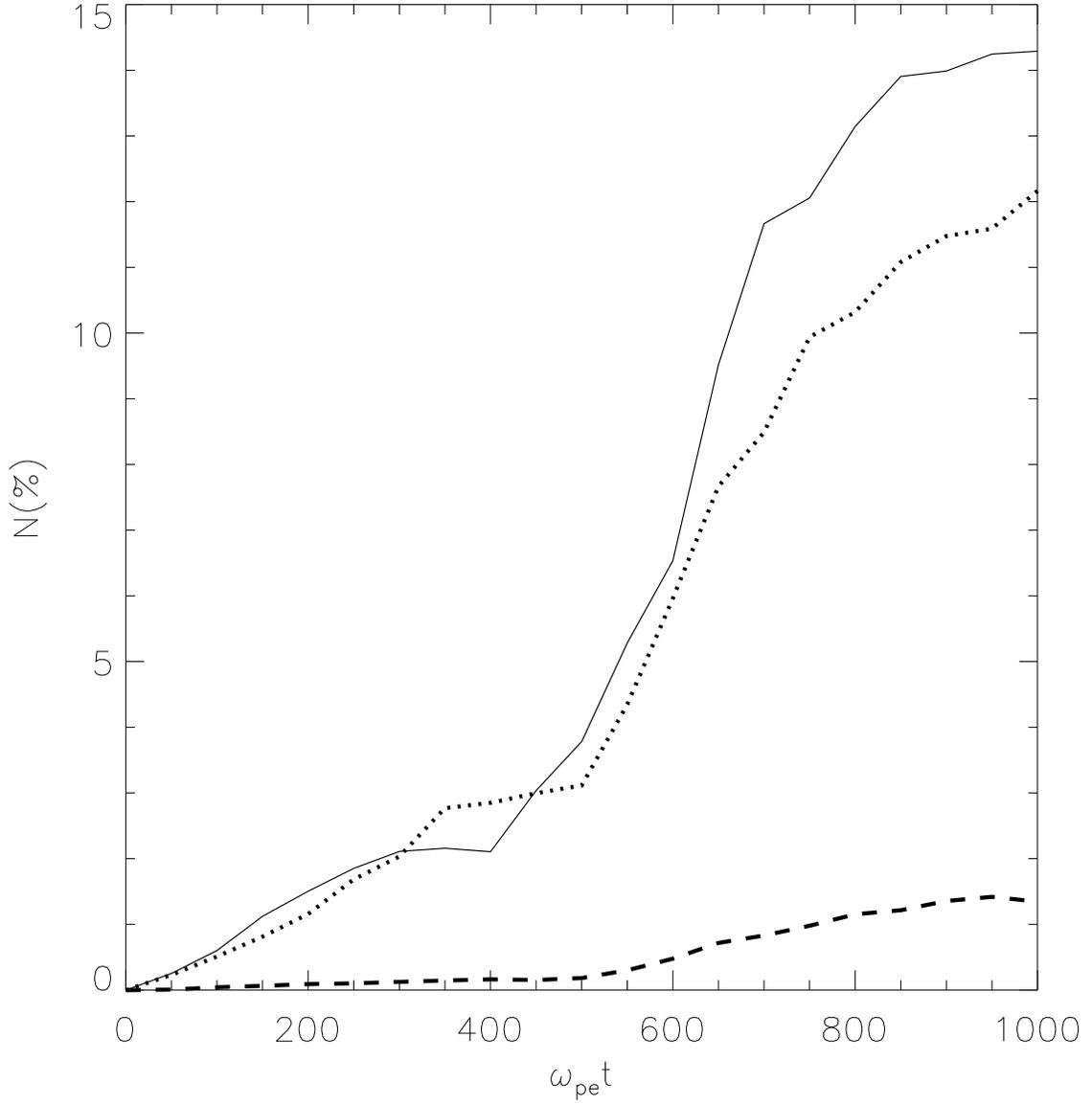}
\caption{The number of accelerated electrons with $E/mc^2$ $>$ 4 expressed
    in the ratio to total electron number for Case I (full line), Case II (dashed line), and
    Case III (dotted line).}
\end{figure}

\begin{figure}
    \epsscale{0.5}
\plotone{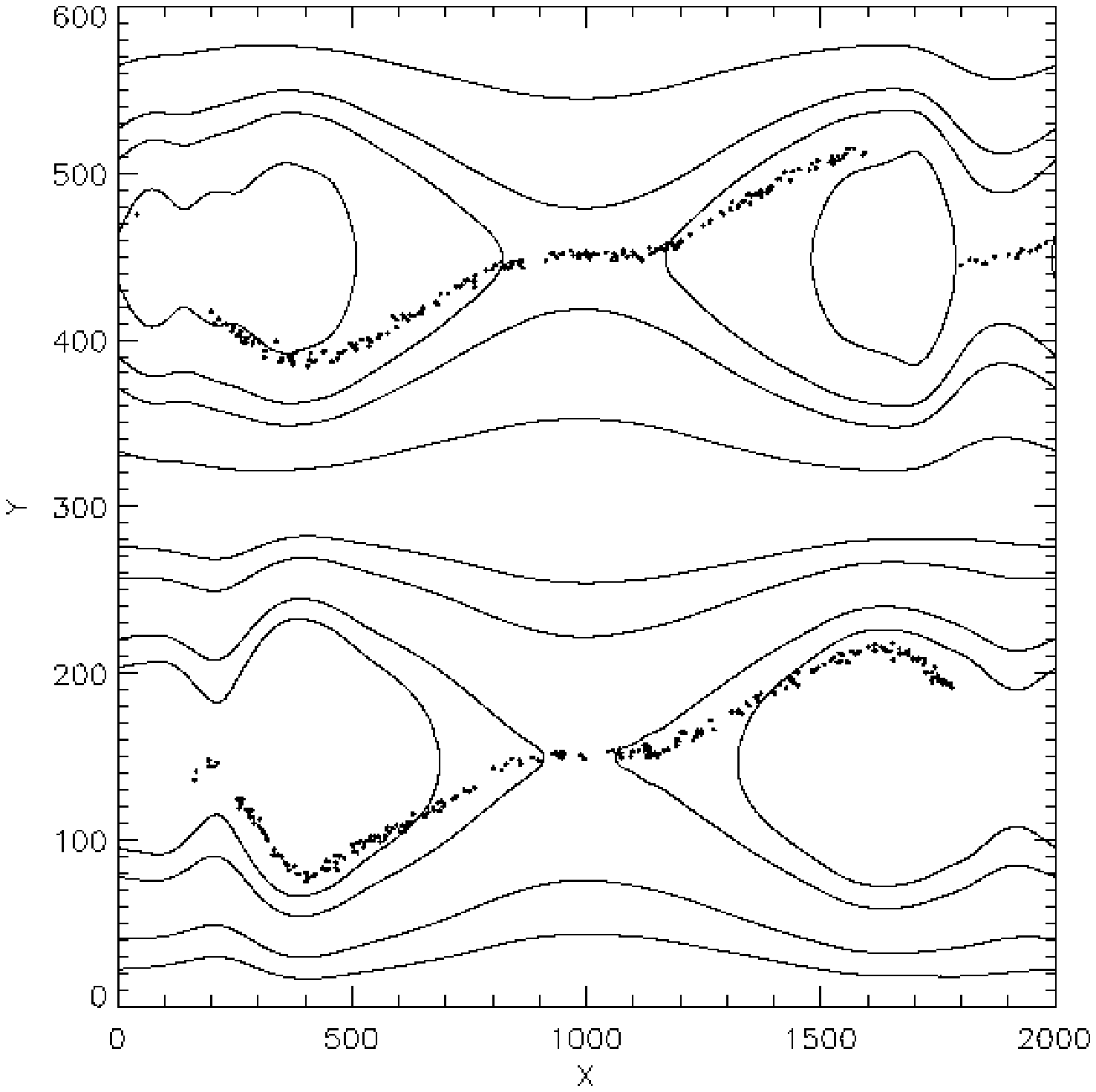} \plotone{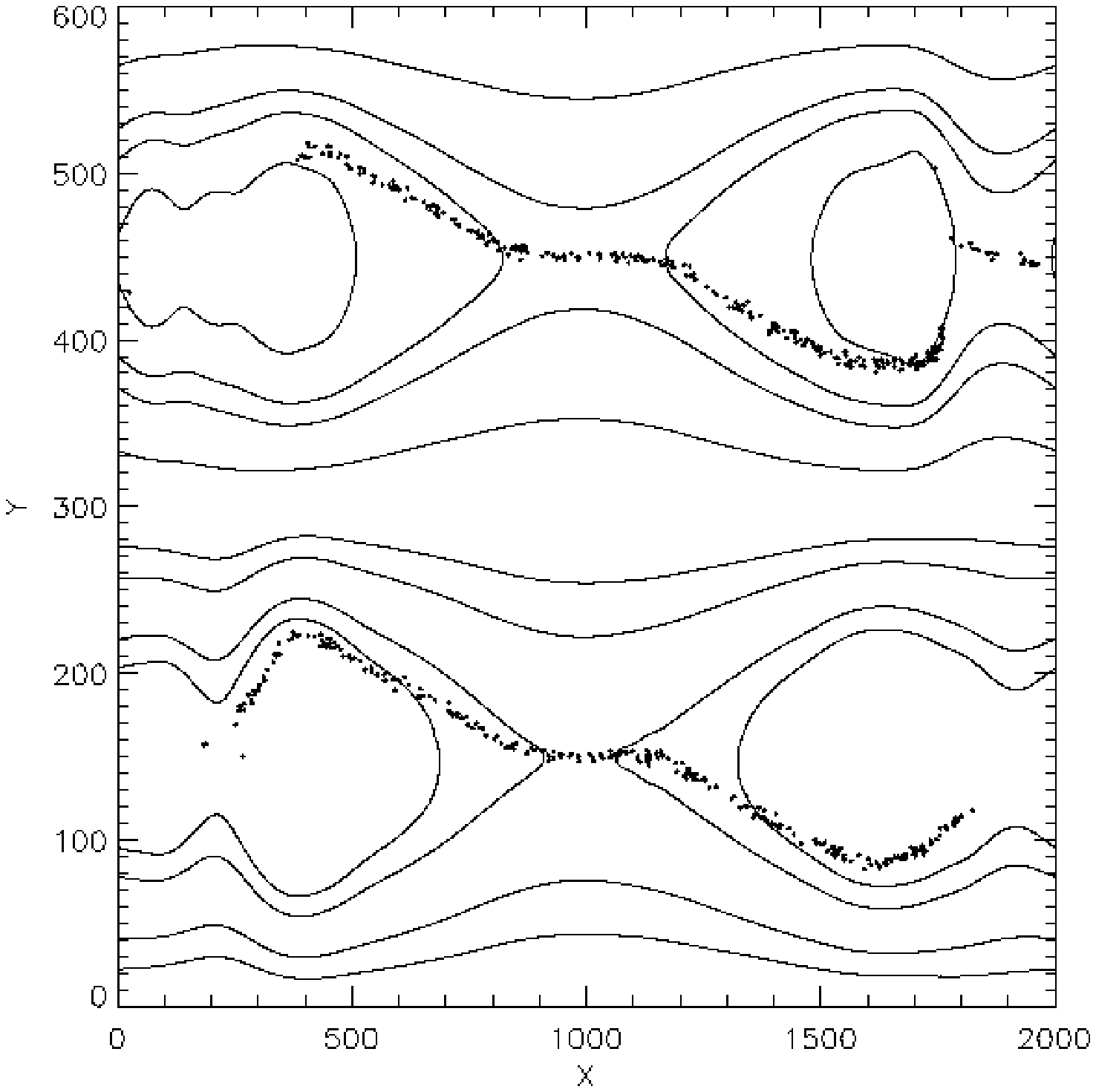}
\caption{The spatial distribution of accelerated electrons (dots in upper part)
    and positrons (dots in bottom
    part) with
    the energy $E/mc^2$ $>$ 10
    superposed on the magnetic field lines projected to the $x-y$ plane
    at $\omega_{pe} t$ = 150 (Case I). Compare locations of electrons and positrons.}
\end{figure}

\begin{figure}
    \epsscale{0.5}
\plotone{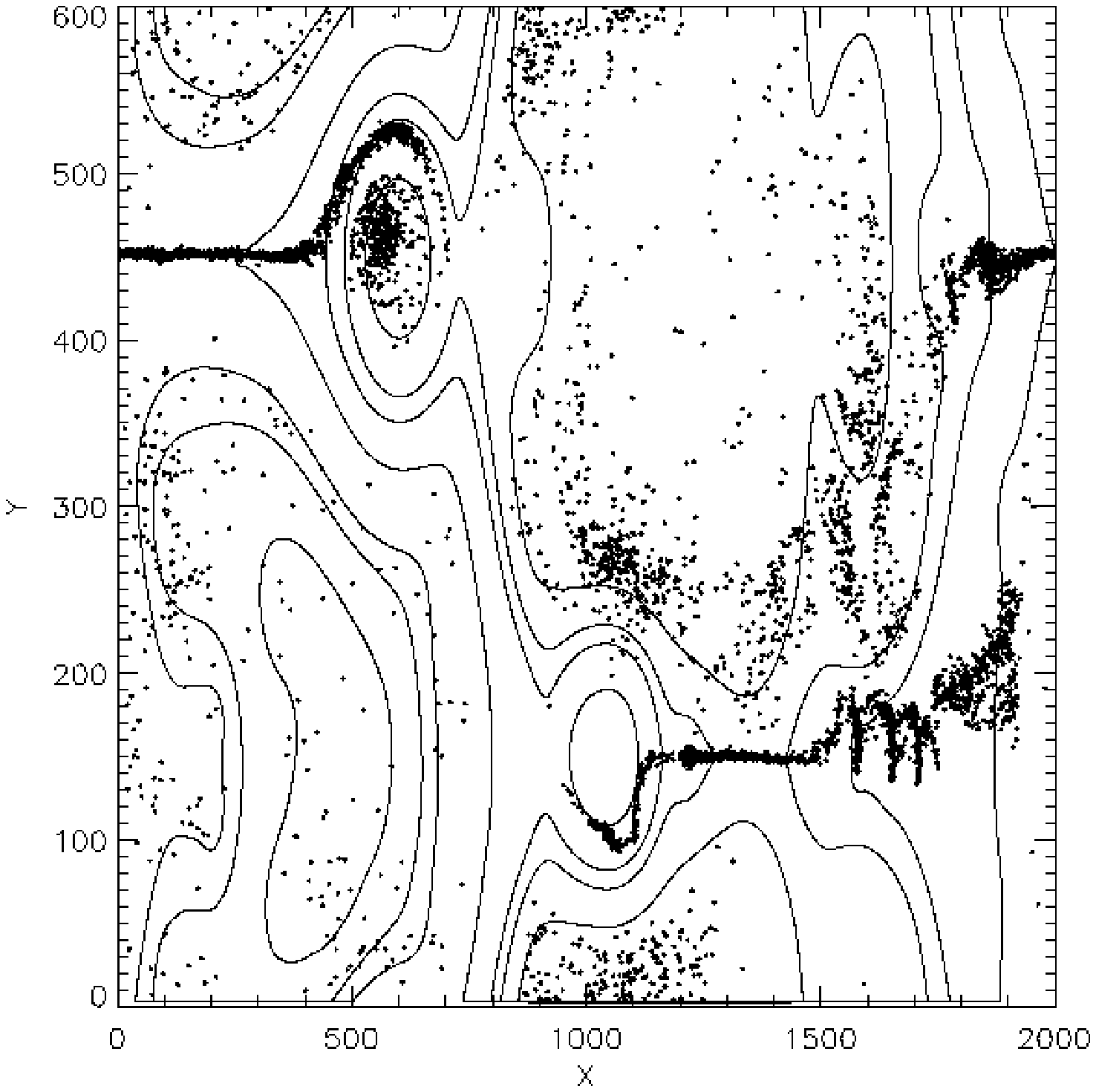} \plotone{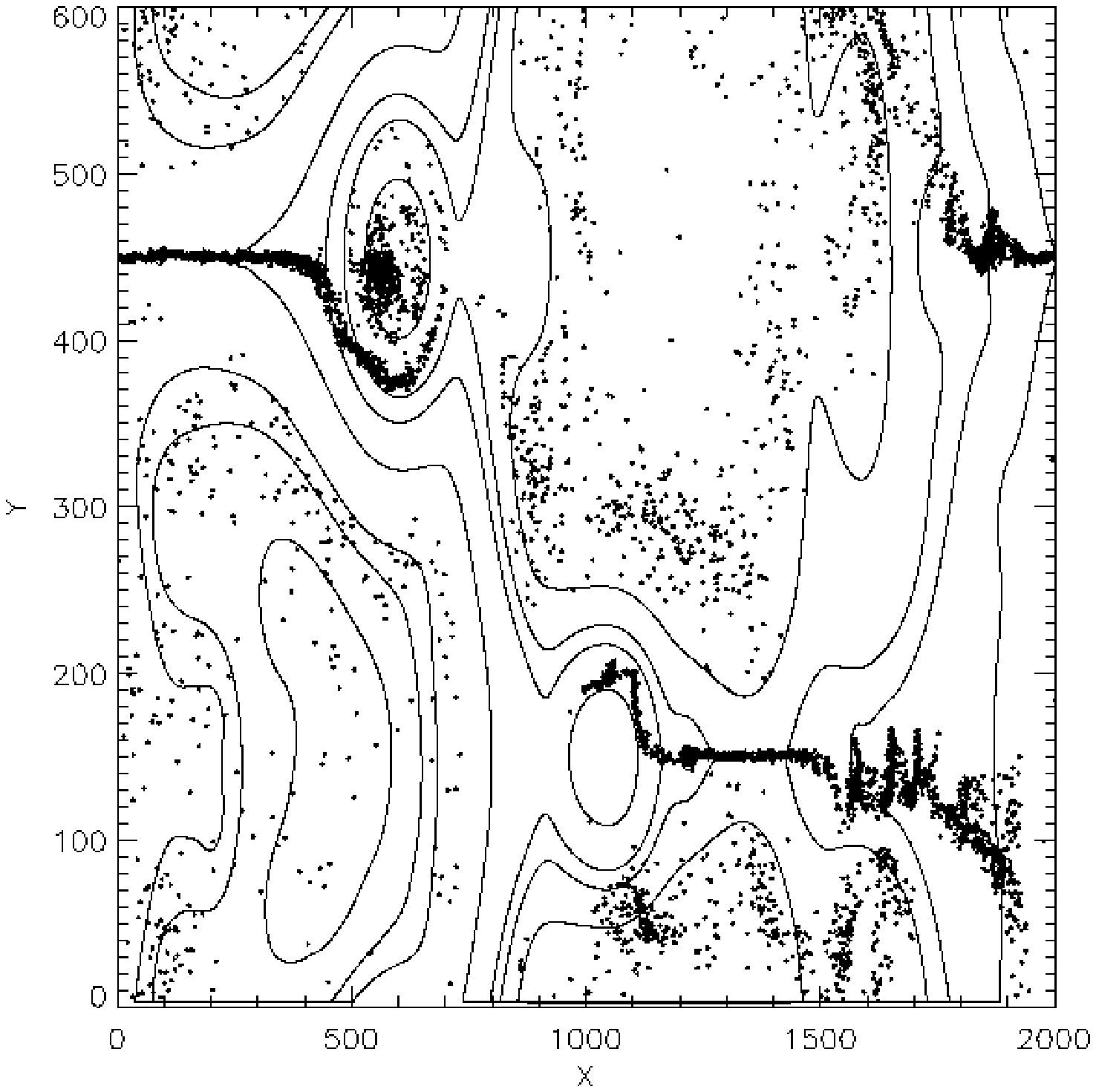}
\caption{The spatial distribution of accelerated electrons (dots in the upper part)
    and positrons (dots in the bottom part) with the energy $E/mc^2$ $>$
    13 superposed on the magnetic field lines projected
    to the $x-y$ plane at $\omega_{pe} t$ = 600 (Case I). For the separation of electrons and
    positrons, see the region around $x$ = 500$\Delta$ and $y$ = 450$\Delta$.}
\end{figure}

\begin{figure}
    \epsscale{1.}
\plotone{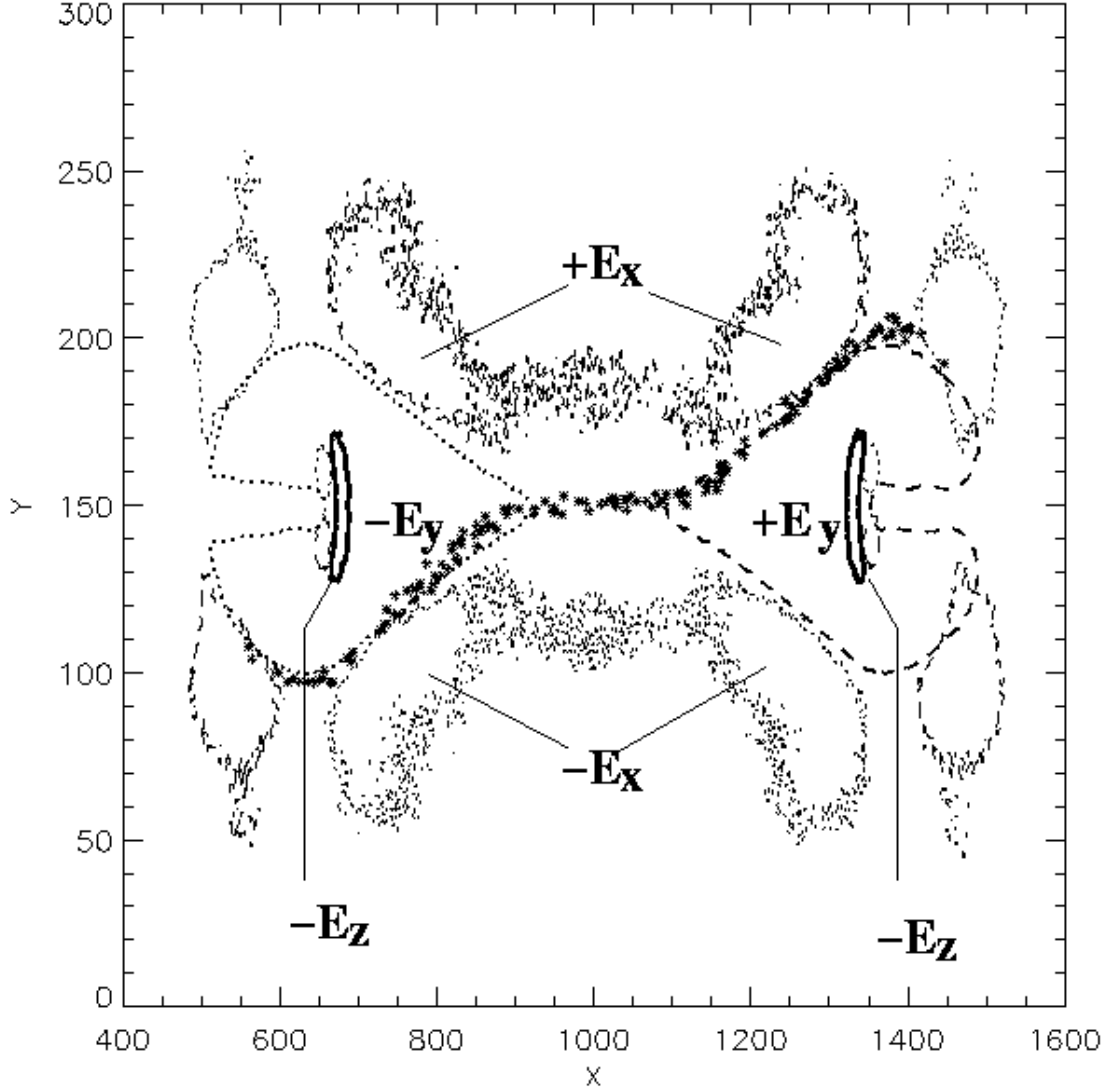}
\caption{The spatial distributions of the electric field components ({\bf E}/B$_0$)
    around the
    X-point in the current sheet near $y$ = 150$\Delta$ at $\omega_{pe} t$ = 100
    (Case I), compare with Figure 6.
    The asterisks mean accelerated electrons for the energy $E/mc^2 >$ 10. The thick full
    contour means the area of enhanced $\mid E_z \mid$ component (level $E_z/B_0$=-0.22).
    (Remark: In the upper current sheet with oppositely oriented current along the line
    $y$=450$\Delta$ the $E_z$ component is oppositely
    oriented, i.e. positive.)
    The thick dashed
    contour is the area of the enhanced $E_y$ (level $E_y/B_0$=0.36). The thick dotted contour
    means the area with oppositely oriented $E_y$ (level $E_y/B_0$=-0.36). The thin dashed
    contour is the area of the enhanced $E_x$ (level $E_x/B_0$=0.07). The thin dotted contour
    means the area with oppositely oriented $E_x$ (level $E_x/B_0$=-0.07).}
\end{figure}

\begin{figure}
    \epsscale{1.}
\plotone{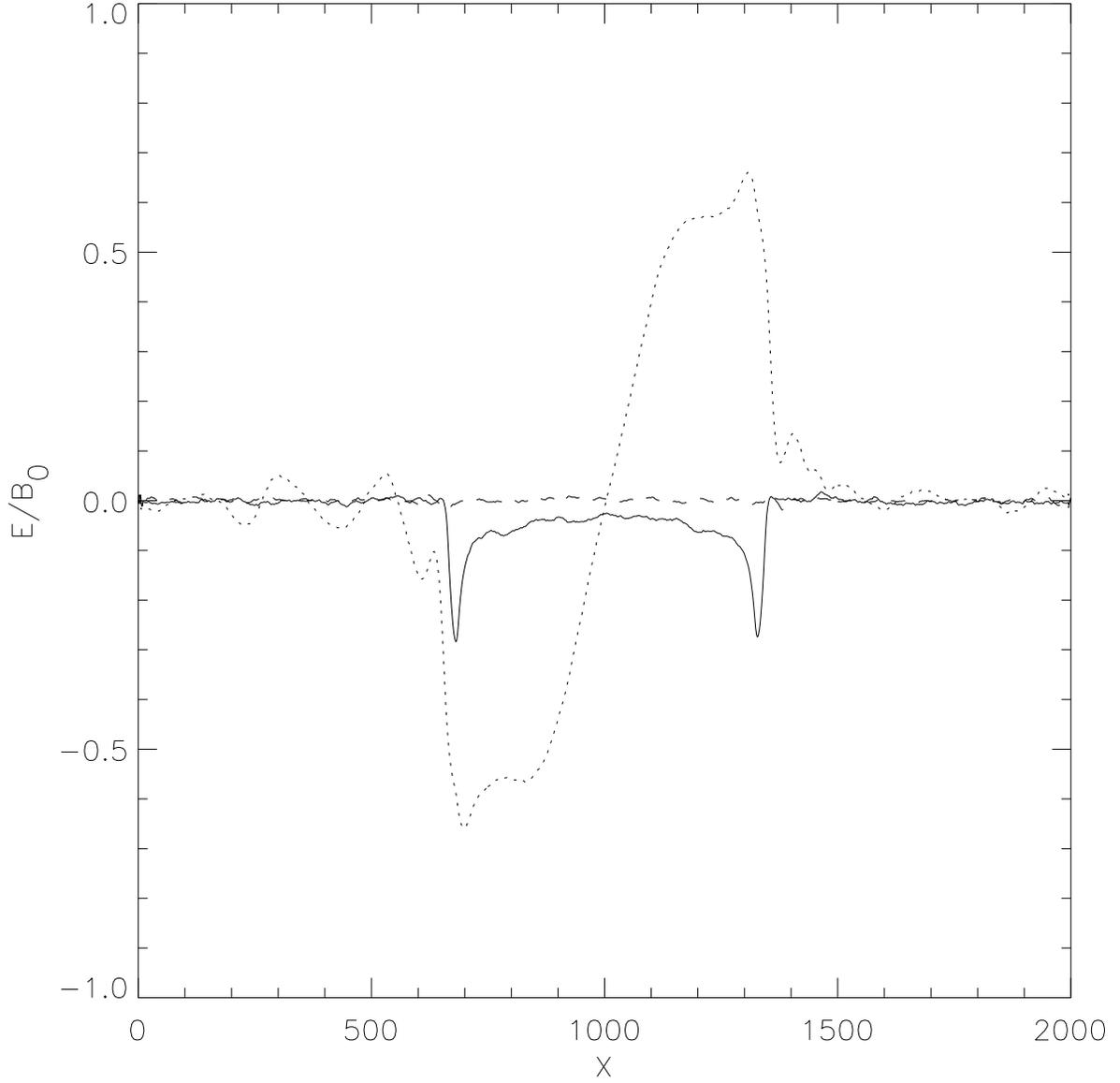}
\caption{The electric field components ($E_x$ (dashed line), $E_y$ (dotted line),
    and $E_z$ (full line)) along the center of the current sheet, i.e. along the line
    $y$=150$\Delta$ at $\omega_{pe} t$ = 100 (Case I) (compare with Figure 5).}
\end{figure}

\begin{figure}
    \epsscale{1.}
\plotone{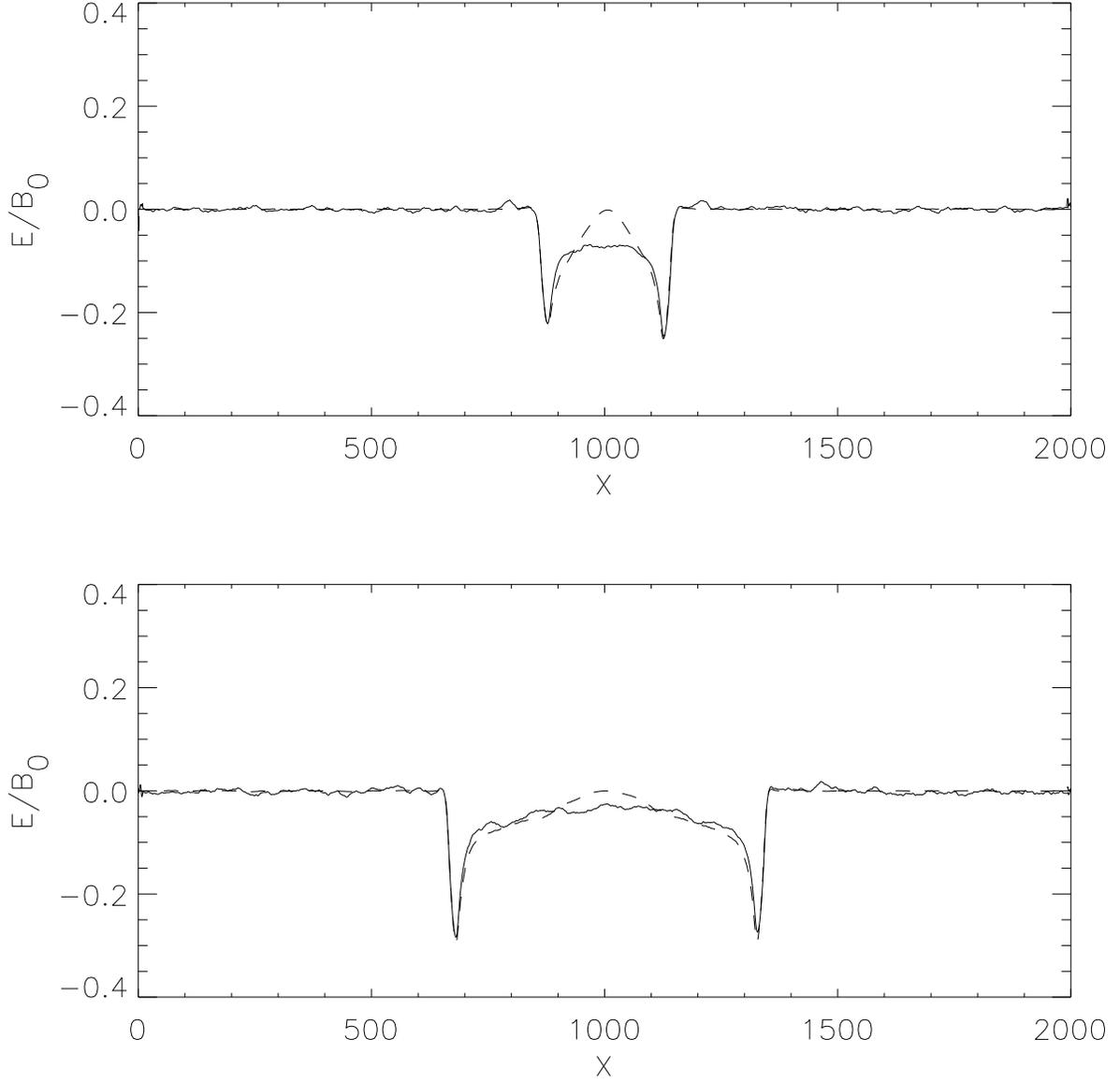}
\caption{The electric field component $E_z$ (full line) and the term
$-{\bf v \times B}/c$ (dashed line) at two times (Case I): at $\omega_{pe} t$ =
50 (upper part), and at $\omega_{pe} t$ = 100 (bottom part).}
\end{figure}

\begin{figure}
    \epsscale{0.5}
\plotone{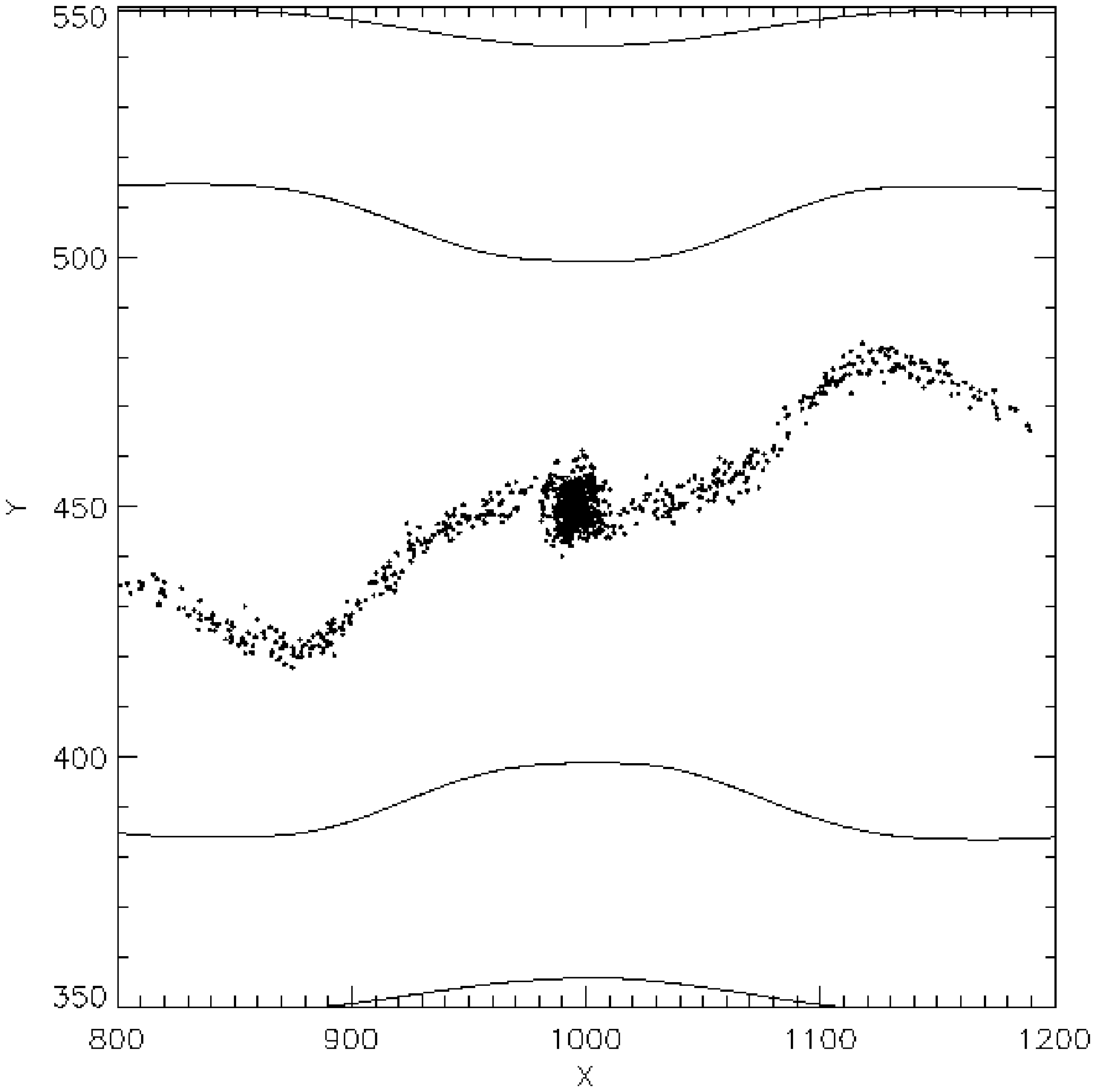} \plotone{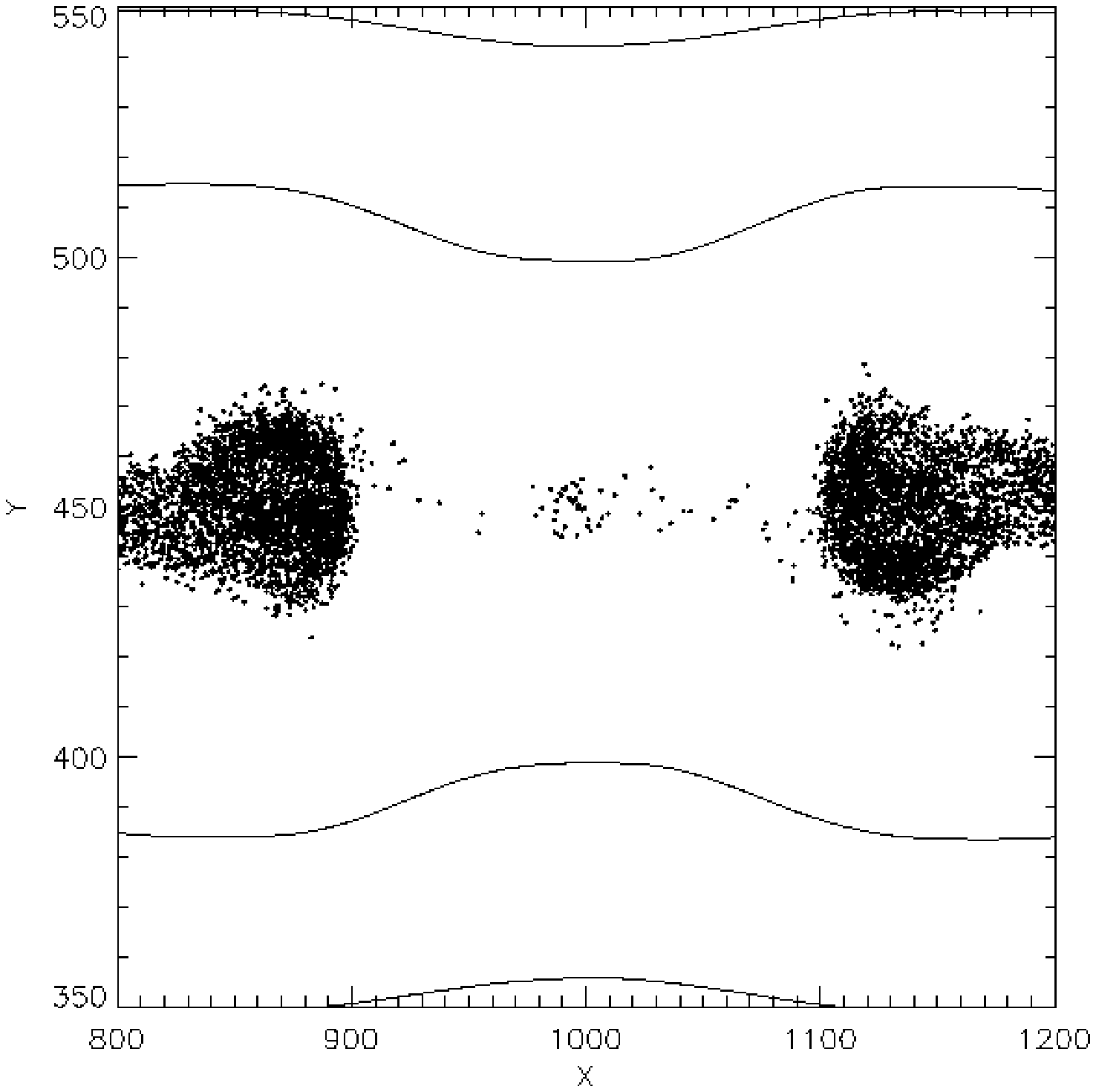}
\caption{The spatial distribution of accelerated electrons (dots in the upper part) with
    the energy $E/m_ec^2$ $>$ 5
    and "protons" (dots in the bottom part) with
    $E/m_ic^2$ $>$ 1.4
    near the X-point of the magnetic field structure in the $x-y$ plane
    at $\omega_{pe} t$ = 50 (Case III). The proton-electron mass ratio is 16.}
\end{figure}


\begin{thebibliography}{}

\bibitem[Bessho \& Bhattacharje (2005)]{BesshoBhattacharje05}
Bessho, N., \& Bhattacharje, A. 2005, Phys. Rev. Letters, 95, 245001

\bibitem[Coronity (1990)]{Coronity90}
Coroniti, F.V. 1990, \apj, 349, 538

\bibitem[Drake et al. (2005)]{Drakeetal2005}
Drake, J.F., Shay, M.A., Thongthai, W., \& Swisdak, M. 2005, Phys. Rev.
Letters, 94, 095001

\bibitem[Drake et al. (2006)]{Drakeetal2006}
Drake, J.F., Swisdak, M., Che, H.,\& Shay, M.A. 2006, Nature, 443/5, 553

\bibitem[Drenkhahn (2002a)]{Drenkhahn02a}
Drenkhahn, G. 2002a, \apj, 387, 714

\bibitem[Drenkhahn (2002b)]{Drenkhahn02b}
Drenkhahn, G. 2002b, \apj, 391, 1141

\bibitem[Giuliani et al. (2005)]{Giulianietal05}
Giuliani, P., Neukirch, T., \& Wood, P. 2005, \apj, 635, 636

\bibitem[Jaroschek et al. (2004a)]{Jaroscheketal04a}
Jaroschek, C.H., Lesch, H.,\& Treumann R.A. 2004a, \apj, 605, L9

\bibitem[Jaroschek et al. (2004b)]{Jaroscheketal04b}
Jaroschek, C.H., Treumann R.A., Lesch, H., \& Scholer, M. 2004b, Physics of
Plasmas, 11 (3), 1151

\bibitem[Karlick\'y \& B\'arta (2006)]{KarlickyBarta06}
Karlick\'y, M., \& B\'arta, M. 2006, \apj, 647, 1472

\bibitem[Karlick\'y \& B\'arta (2007)]{KarlickyBarta07}
Karlick\'y, M., \& B\'arta, M. 2007, \aap, 464, 735

\bibitem[Larrabee et al. (2003)]{Larrabee03}
Larrabee, D., Lovelace, R., \& Romanova, M. 2003, \apj, 586, 72

\bibitem[Lesch \& Birk (1998)]{LeschBirk98}
Lesch, H., \& Birk, G.T. 1998, \apj, 499, 167

\bibitem[Litvinenko (1996)]{Litvinenko96}
Litvinenko, Yu., E. 1996, \apj, 462, 997

\bibitem[Lyubarsky \& Kirk (2001)]{LyubarskyKirk01}
Lyubarsky, Y., \& Kirk, J. 2001, \apj, 547, 437

\bibitem[Martens \& Young (1990)]{MartensYoung90}
Martens, P.C.H., \& Young, A. 1990, \apjs, 73, 333

\bibitem[Michel (1994)]{Michel994}
Michel, F. 1994, \apj, 431, 397

\bibitem[Priest \& Forbes (2000)]{PriestForbes2000}
Priest, E.R.,\& Forbes, T. 2000, Magnetic Reconnection: MHD Theory and
Applications, Cambridge Univ. Press, Cambridge, UK

\bibitem[Pritchett (2006)]{Pritchett2006}
Pritchett, P.L. 2006, J. Geophys. Res., 111, A10212

\bibitem[Saito \& Sakai (2004)]{SaitoSakai2004}
Saito, S., \& Sakai, J.I. 2004, \apj, 616, L179

\bibitem[Share et al. (3003)]{Shareetal2003} Share, G.H., Murphy, R.J., Smith, D.M., Lin, R.P., Dennis, B.R.,
\& Schwartz, R.A. 2003, \apj, 595, L89

\bibitem[Wardle et al. (1998)]{Wardleetal98}
Wardle, J., Homan, C., Ojiha, R., \& Roberts, D. 1998, Nature, 395, 457

\bibitem[Zenitani \& Hoshino (2001)]{ZenitaniHoshino01}
Zenitani, S., \& Hoshino, M. 2001, \apj, 562, L63

\bibitem[Zenitani \& Hoshino (2005)]{ZenitaniHoshino05}
Zenitani, S., \& Hoshino, M. 2005, \apj, 618, L111

\bibitem[Zharkova \& Gordovskyy]{Zharkova2004}
Zharkova, V.V.,\& Gordovskyy, M. 2004, \apj, 604, 884

\bibitem[Zhu \& Parks (1993)]{ZhuParks93}
Zhu, Zh., \& Parks, G. 1993, J. Geophys. Res., 98, 7603




\end{thebibliography}
\end{document}